\documentclass[twocolumn,amsmath,amssymb]{revtex4}

\usepackage{psfig}

\begin{document}

\title{Two-loop QCD corrections to top quark decay}

\author{Maciej \'Slusarczyk\footnote{Work based on collaboration with
I.~Blokland, A.~Czarnecki and F.~Tkachov. A contribution to the
2004 Lake Louise Winter Institute proceedings.}}

\affiliation{
Department of Physics, University of Alberta\\
Edmonton, AB\ \  T6G 2J1, Canada \\
and \\
Institute of Physics,  Jagiellonian University, \\
Reymonta 4, Krak\'{o}w, Poland }

\begin{abstract}
We present a determination of a new class of Feynman diagrams
relevant for second-order QCD corrections to the top quark decay
$t \rightarrow b W$. Modern computing techniques allow us to
perform a reduction of the original loop integrals to master
integrals. We obtain the analytical value of the top decay rate as
an expansion around the limit of massless $b$ and $W$.
\end{abstract}


\maketitle

%%%%%%%%%%%%%%%%%%%%%%%%%%%%%%%%%%%%%%%%%%%%%%%%%%%%%%%%%%%%%%%%%%%%%%%%%%%%%%%%%%%%%%
\section{Introduction}
%%%%%%%%%%%%%%%%%%%%%%%%%%%%%%%%%%%%%%%%%%%%%%%%%%%%%%%%%%%%%%%%%%%%%%%%%%%%%%%%%%%%%%
It is well-known that the top quark is the heaviest fermion in the
Standard Model. An immediate consequence of that fact is its very
short lifetime which turns out to be shorter than the
characteristic timescale of non-perturbative QCD effects. Thanks
to this feature, the top behaves almost like a free quark, which
is really a unique situation in QCD. Moreover, the process
$t\rightarrow bW$ is an absolutely dominant decay channel.
Additional decay channels arise in various extensions of the
present Standard Model (like $t \rightarrow b H$ or $ t
\rightarrow $ supersymmetric particles). For these reasons, top
physics seems to be a particularly clean laboratory for new
physics searches but two complementary pieces of information are
necessary. First, we need an accurate experimental determination
of the top decay rate, which will be feasible in the future, e.g.
at the Next Linear Collider.  Also, to fully utilize this
measurement it is incredibly important to have theoretical
predictions in the framework of the present Standard Model with
accuracy matching the experimental precision.

Much effort has been invested into studies of radiative
corrections to the top quark decay width. The one-loop QCD
contribution was computed in \cite{Jezabek}. It decreases the
tree-level rate by about $8.4 \%$. The electroweak NLO part was
evaluated in \cite{electroweak} but its effect turned out to be
much smaller --- it increases the decay rate by about $2 \%$.
Moreover, the electroweak contribution is almost entirely canceled
when one takes into account the finite $W$ width \cite{Jezabek2}.
The NNLO QCD contribution has been studied so far in various
kinematical limits. For example, there exists a result in zero
recoil kinematics with $m_W$ set to zero \cite{Czarnecki1}, as
well as a numerical study done by means of Pad\'e approximations
\cite{Chetyrkin}. All these calculations estimate the NNLO QCD
contribution at about $-2 \%$ but, for comparison with planned
measurements, a precise and preferably analytical result would be
of interest.

In the present study we start with the limit of a massive top
quark decaying into a massless $b$ quark and a massless $W$ boson.
Then we construct an expansion in the artificial parameter
$(m_W/m_t)^2$. Taking sufficiently many expansion terms, we can
closely approach the physical point $(m_W/m_t)^2 \simeq 0.213$ and
give a precise prediction for the second order QCD contribution to
the decay rate.

%%%%%%%%%%%%%%%%%%%%%%%%%%%%%%%%%%%%%%%%%%%%%%%%%%%%%%%%%%%%%%%%%%%%%%%%%%%%%%%%%%%%%%
\section{Methods used in the calculation}
%%%%%%%%%%%%%%%%%%%%%%%%%%%%%%%%%%%%%%%%%%%%%%%%%%%%%%%%%%%%%%%%%%%%%%%%%%%%%%%%%%%%%%
Let us now briefly discuss the computing methods which were used
in the calculation. In the traditional approach the contribution
to the decay rate can be divided into three distinct classes: real
radiation with gluons in the final state, virtual loop
corrections, as well as a mixed real-virtual case. Effective tools
for computing multi-loop diagrams have been developed over the
last few years. Conversely, real radiation requires integration
over phase space in the final state, which has to be done manually
on a case-by-case basis. This usually constitutes the bottleneck
of the computation and an automated method for dealing with real
radiation seems mandatory at NNLO.

\begin{figure}[htb]
\begin{tabular}{cc}
\psfig{figure=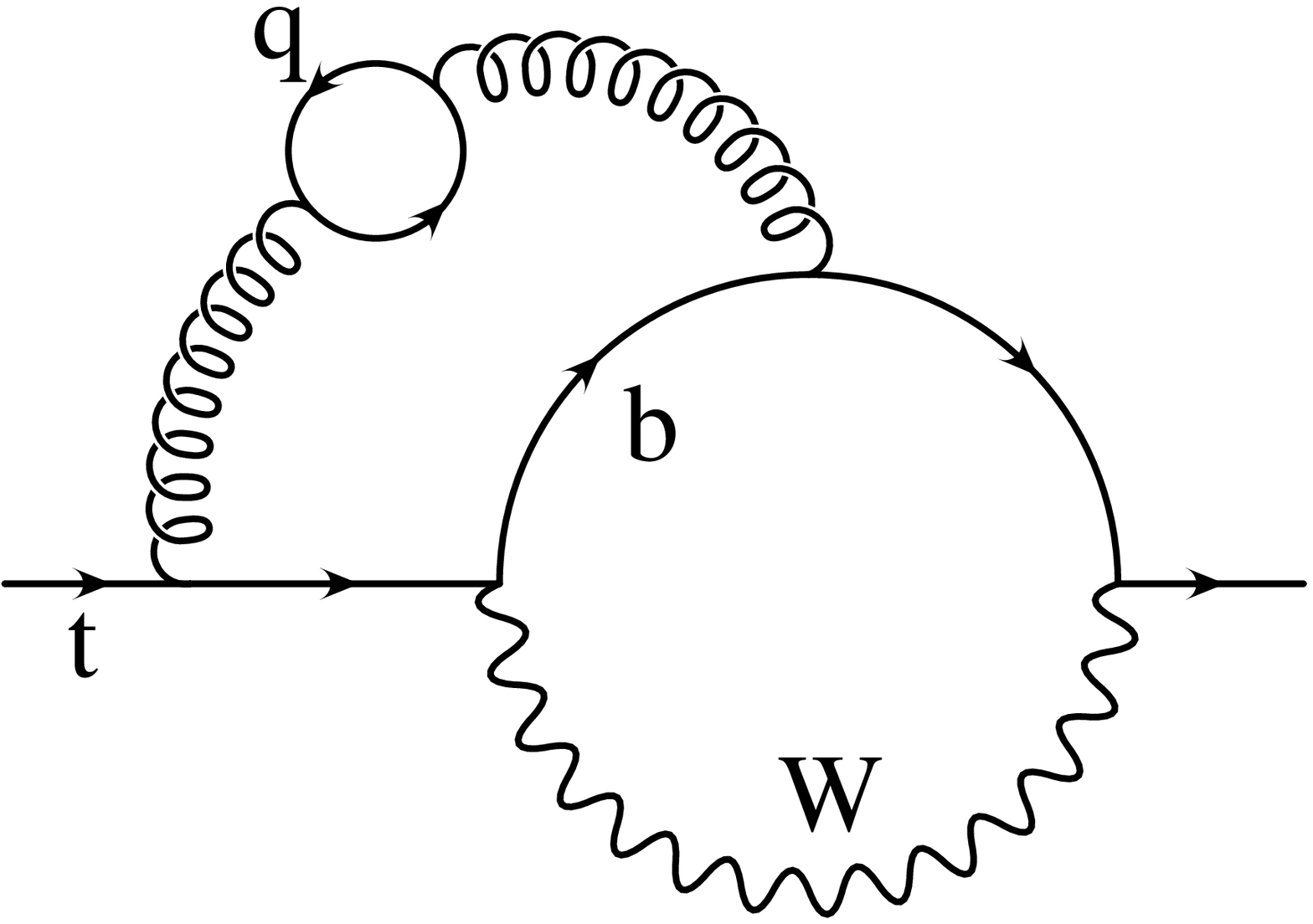,width=39mm} & \hspace{5mm}
\psfig{figure=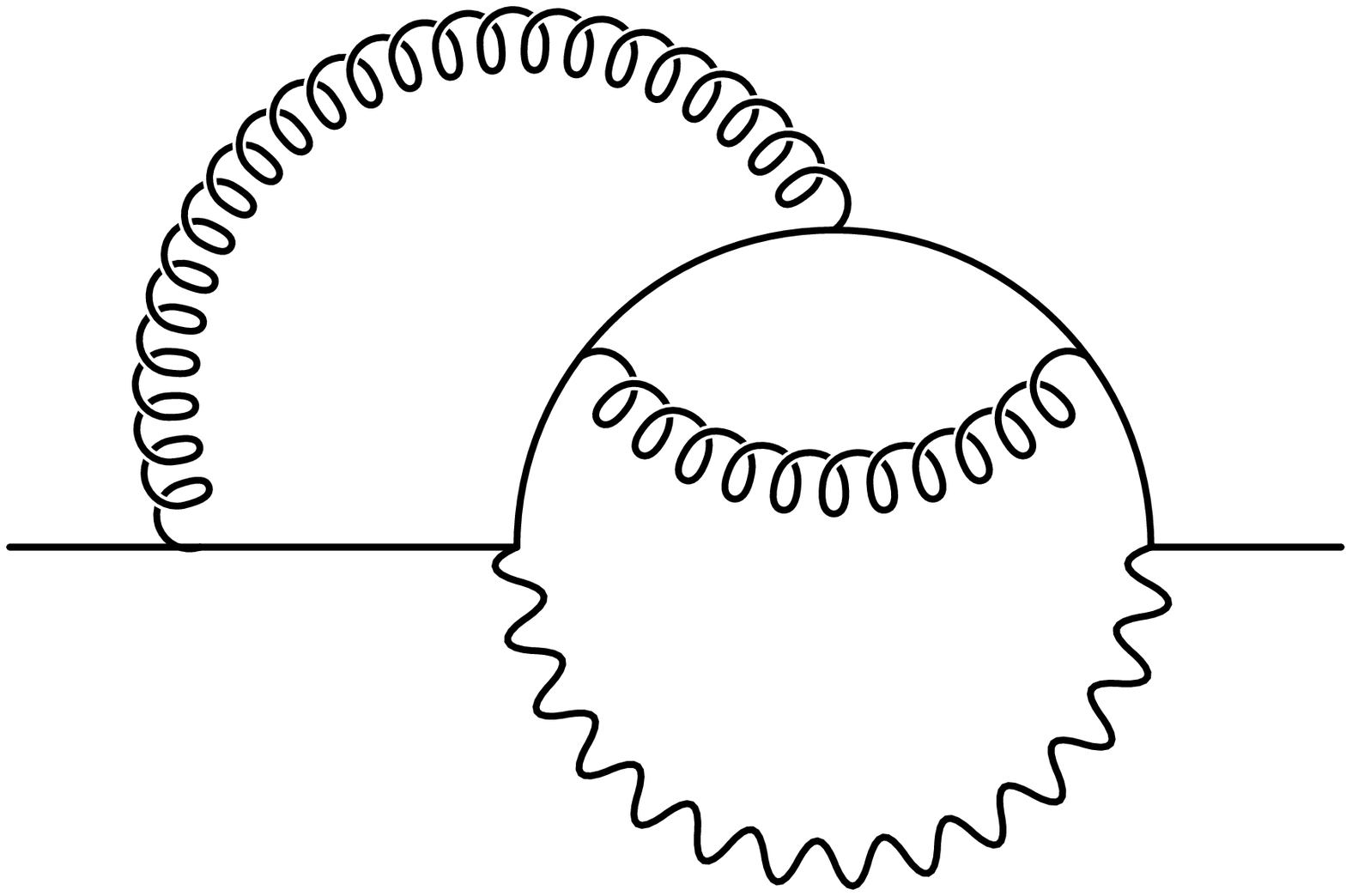,width=39mm} \\
(a) & (b)
\end{tabular}
\begin{tabular}{c}
\psfig{figure=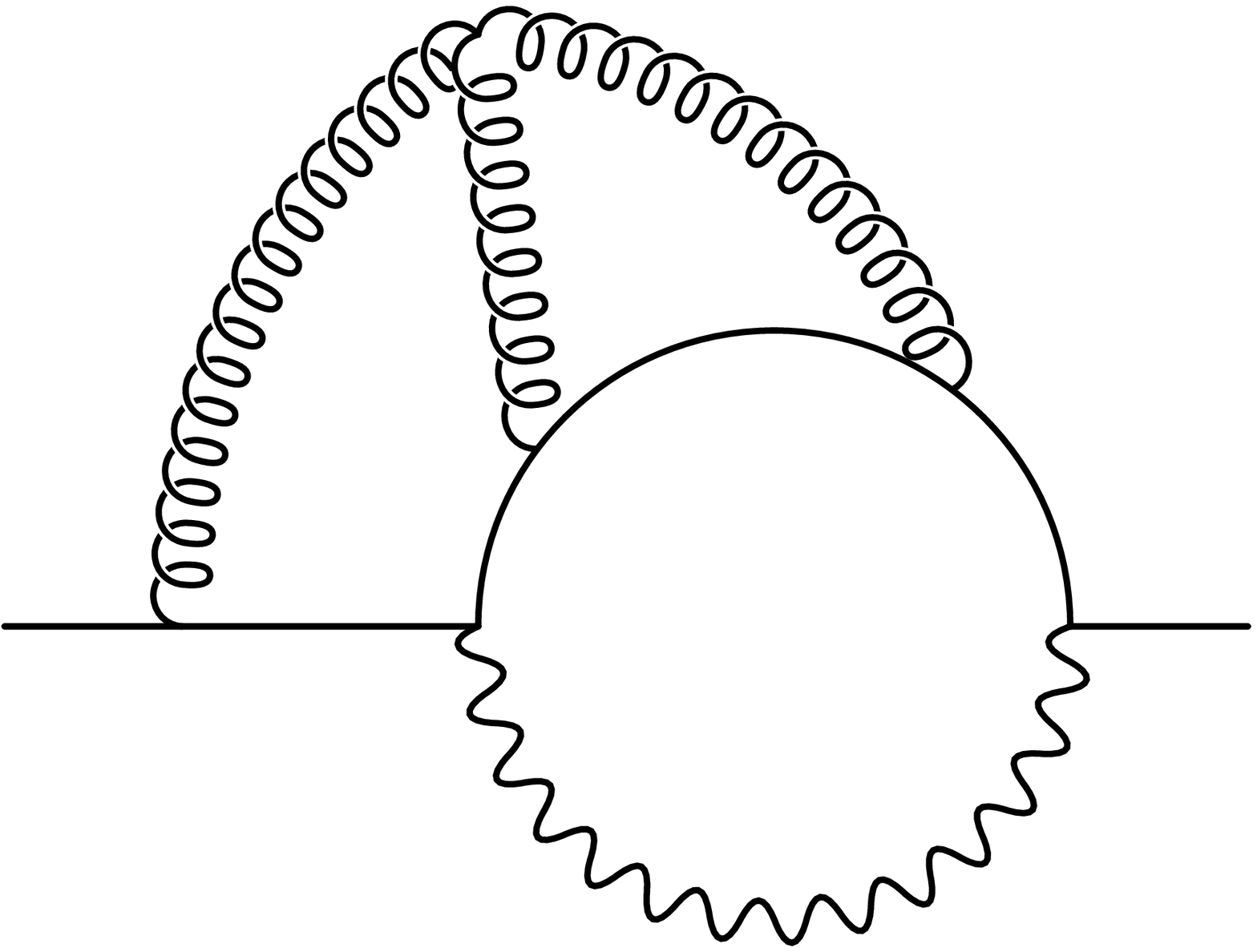,width=39mm} \\
(c)
\end{tabular}
\caption{Examples of diagrams whose cuts contribute to the
$\mathcal{O}(\alpha_s^2)$ decay rate for $t\rightarrow bW$: (a)
light or heavy quarks; (b) abelian; (c) non-abelian.}
\label{fig:diagrams}
\end{figure}

A very simple but powerful idea is to map all real radiation
diagrams onto loop integrals. To achieve this goal we apply the
optical theorem, which relates contributions to the decay to
imaginary parts of self-energy diagrams. In
Fig.~\ref{fig:diagrams} we show three examples of the diagrams
that we have to consider in order to calculate $t\rightarrow bW$
at $\mathcal{O}(\alpha_s^2)$.  There are $19$ abelian diagrams,
$11$ non-abelian diagrams, and $6$ diagrams with quark vacuum
polarization loops that have to be taken into account. The various
cuts of these diagrams correspond to two-loop virtual corrections
or emissions of one or two real quanta. However, the price to pay
for the automation gained by expressing these diagrams in terms of
a uniform set of self-energy graphs is the necessity of computing
the imaginary parts of three-loop diagrams.
This difficult task can be organized in the following way. First,
we perform a reduction of the original integrals to a set of
simpler master integrals. In this phase we use the so-called
integration-by-parts identities\cite{Fyodor} which lead to a
system of recurrence relations between various loop integrals. The
solution of such a system provides reduction formulas which allow
us to express the original loop integrals in terms of master
integrals.

There are two main paradigms for solving the system of recurrence
relations. In the traditional method, one inspects the structure
of the identities and rearranges them manually for an efficient
iterative solution of the system. This ``by inspection" method has
proven to be very successful in numerous applications (see, e.g.,
\cite{vanRittbergen1} -- \cite{Broadhurst}) but requires much
human effort to implement. A newer method is based on the Gaussian
elimination of a large system of identities with a given ordering
function which measures the ``difficulty" of each integral
\cite{Laporta}. This approach is fully automated and
process-independent, but requires significant computing power.
Thanks to an effective implementation and the growing performance
of hardware, this method has become feasible at NNLO quite
recently. In the present project we used both methods in parallel.
We obtained identical results which serve as a crosscheck for the
calculation. It also allows us to compare the two approaches and
point out strengths and weaknesses of both.

The final step is the evaluation of the master integrals arising
in the problem. For $t\rightarrow bW$ at $\mathcal{O}(\alpha_s^2)$
there are $23$ master integrals which have been computed
analytically.

%%%%%%%%%%%%%%%%%%%%%%%%%%%%%%%%%%%%%%%%%%%%%%%%%%%%%%%%%%%%%%%%%%%%%%%%%%%%%%%%%%%%%%
\section{Results}
%%%%%%%%%%%%%%%%%%%%%%%%%%%%%%%%%%%%%%%%%%%%%%%%%%%%%%%%%%%%%%%%%%%%%%%%%%%%%%%%%%%%%%
Taking advantage of the tools discussed in the previous section,
we can obtain the decay rate in the form of an expansion around
the limit $m_W = m_b = 0$ in the parameter $\omega = (m_W/m_t)^2$
(see \cite{toppaper}). The decay width can be written in the
following form:
\begin{equation}
\Gamma(t\rightarrow bW) = \Gamma_0 \left[ X_0 +
\frac{\alpha_s}{\pi} X_1 + \left( \frac{\alpha_s}{\pi} \right)^2
X_2 \right],
\end{equation}
where
\begin{equation}
\Gamma_0 \equiv \frac{G_F m_t^3 \left| V_{tb}
\right|^2}{8\sqrt{2}\pi},
\end{equation}
$X_0$ corresponds to the Born decay rate and $X_1$ is the already
known NLO correction. Our goal in the present calculation is the
coefficient $X_2$, which can be subdivided into four
gauge-invariant color pieces:
\begin{equation}
X_2 = C_F \left( T_R N_L X_L + T_R N_H X_H + C_F X_A + C_A X_{NA}
\right),
\end{equation}
where $C_F=4/3$, $C_A=3$, and $T_R=1/2$ are the usual SU(3) color
factors and $N_L$ and $N_H$ denote the number of light $(m_q=0)$
and heavy $(m_q=m_t)$ quark flavors.

\begin{figure}[htb]
\psfig{figure=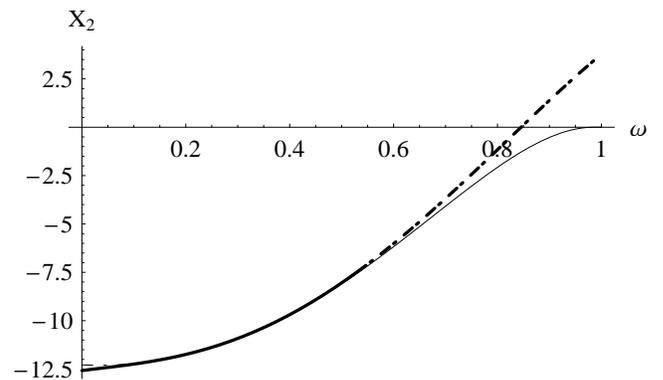,width=85mm} \caption{Matching of expansions
around $\omega = 0$ (thick line) and $\omega = 1$ (thin line). The
solid line denotes the resulting decay width valid in the full
range of $\omega$.} \label{fig:expansion}
\end{figure}

We obtained series coefficients to $\omega^5$ analytically. In
terms of numerical values our expansion reads:
\begin{equation}
X_2 = -12.579 + 3.743 \, \omega + 6.054 \,  \omega^2 +5.619 \,
\omega^2 \log \omega + \mathcal{O}(\omega^3).
\end{equation}
It is a straightforward task to improve the accuracy of an
expansion by computing more terms in $\omega$. Moreover, the
present expansion can be smoothly matched with the one in the zero
recoil kinematics studied previously \cite{Czarnecki} in the
context of semileptonic b quark decays. The result of such a
matching procedure is depicted in Fig.~\ref{fig:expansion}.

For the measured ratio of the $W$ and top masses \cite{pdg},
$\omega \simeq 0.213$, our expansion gives $X_2 = -15.5(1)$, where
the uncertainty is almost entirely due to the experimental error
in the determination of $m_t$. The theoretical error, which
originates from taking a finite number of terms in our expansion,
is $20$ times smaller and can be reduced further if needed. Using
$\alpha_s ( m_t) = 0.11$, we find that the two-loop correction
decreases the tree level decay rate by about $2 \%$, in agreement
with earlier expectations.

%%%%%%%%%%%%%%%%%%%%%%%%%%%%%%%%%%%%%%%%%%%%%%%%%%%%%%%%%%%%%%%%%%%%
\section{\bf{Conclusions}}
%%%%%%%%%%%%%%%%%%%%%%%%%%%%%%%%%%%%%%%%%%%%%%%%%%%%%%%%%%%%%%%%%%%%
We have presented a new analytic $\mathcal{O}(\alpha_s^2)$ result
for the decay $t\rightarrow bW$ in terms of a parameter
$\omega=(m_W/m_t)^2$ and in the limit of $m_b=0$.  This result has
enabled us to confirm or modify slightly the corresponding results
of previous numerical calculations.  Our formulas are readily
applicable to other physical processes such as muon decay and the
semileptonic $b$ quark decay $b\rightarrow u l\nu$.

Our results depend on the imaginary parts of a novel class of
three-loop integrals, which we have obtained using two independent
paradigms for the solution of large systems of recurrence
relations.  To the best of our knowledge, this is the first time
that both approaches have been used simultaneously to obtain a new
result, and an objective analysis of the strengths and weaknesses
of each approach will increase the efficiency of other large
calculations in the future.

\emph{Acknowledgements:} I am grateful to F.~Tkachov for fruitful
collaboration in developing the dedicated computer algebra system
used in this project as well as to I.~Blokland and A.~Czarnecki
for reading the manuscript and helpful remarks. This research was
supported by the Alberta Ingenuity and by the Natural Sciences and
Engineering Research Council of Canada.


\begin{thebibliography}{0}
\bibitem{Jezabek} M. Jezabek and J. Kuhn, {\it Nucl. Phys.}
{\bf B34}, 1729 (1980).
\bibitem{electroweak} A. Denner and T. Sack, {\it Nucl. Phys.}
{\bf B358}, 46 (1991). G. Eilam, R. Mendel, R. Migneron, and A.
Soni, {\it Phys. Rev. Lett.} {\bf 66}, 3105 (1991).
\bibitem{Jezabek2} M. Jezabek and J. Kuhn, {\it Phys. Rev.}
{\bf D48}, 1910 (1993).
\bibitem{Czarnecki1} A. Czarnecki and K. Melnikov, {\it Nucl.
Phys.} {\bf B544}, 520 (1999).
\bibitem{Chetyrkin} K.~G. Chetyrkin, R. Harlander, T.
Seidensticker, and M. Steinhauser, {\it Phys. Rev.} {\bf D60},
114015 (1999).
\bibitem{Fyodor} F. V. Tkachov, {\it Phys. Lett.} {\bf B100}, 65
(1981).
\bibitem{vanRittbergen1} T. van Ritbergen and R.~G. Stuart,
{\it Phys. Rev. Lett.} {\bf 82}, 488 (1999).
\bibitem{vanRittbergen2} T. van Ritbergen, {\it Phys. Lett.}
{\bf B454}, 353 (1999).
\bibitem{Czarnecki} A. Czarnecki and K. Melnikov, {\it Phys. Rev. Lett.}
{\bf 88}, 131801 (2002).
\bibitem{Broadhurst} D.~J. Broadhurst, {\it Z. Phys.}
{\bf C54}, 599 (1992).
\bibitem{Laporta} S. Laporta, {\it Int. J. Mod. Phys.}
{\bf A15}, 5087 (2000).
\bibitem{toppaper} I. Blokland, A. Czarnecki, M. Slusarczyk, and F. Tkachov,
hep-ph/0403221.
\bibitem{pdg} K. Hagiwara {\it et al.}, {\it Phys. Rev.}, {\bf
D66}, 010001 (2002).
\end{thebibliography}
\end{document}